# Generation and annotation of item usage scenarios in e-commerce using large language models

*Madoka Hagiri, Kazushi Okamoto, Koki Karube, Kei Harada, and Atsushi Shibata
The University of Electro-Communications, Tokyo, Japan
{m-hagiri, kazushi, karubekoki, harada, shibata-atsushi}@uec.ac.jp

**Abstract**

Complementary recommendations suggest combinations of useful items that play important roles in e-commerce. However, complementary relationships are often subjective and vary among individuals, making them difficult to infer from historical data. Unlike conventional history-based methods that rely on statistical co-occurrence, we focus on the underlying usage context that motivates item combinations. We hypothesized that people select complementary items by imagining specific usage scenarios and identifying the needs in such situations. Based on this idea, we explored the use of large language models (LLMs) to generate item usage scenarios as a starting point for constructing complementary recommendation systems. First, we evaluated the plausibility of LLM-generated scenarios through manual annotation. The results demonstrated that approximately 85% of the generated scenarios were determined to be plausible, suggesting that LLMs can effectively generate realistic item usage scenarios.

## 1 Introduction

In e-commerce, recommender systems are widely used to improve user convenience and the purchasing experience. Traditional systems suggest items based on inferred past behavior; however, these do not always match the user's current needs. User intent often changes with the context or purchase stage. For instance, after buying a durable item such as a camera, recommending similar products is redundant. Complementary recommendations address this issue by suggesting items that are typically used together to enhance the overall experience [1]. However, most studies rely on historical data such as purchase histories and behavioral logs, which are often sparse, noisy, and insufficient for capturing subjective or content-dependent relationships [2, 3, 4].

This study proposes a framework inspired by how people identify complementary items. We hypothesize that people first imagine a specific usage scenario, which then provides a context for identifying the necessary items. For instance, a user with a camera might imagine "taking commemorative photos on a family trip," leading to need an SD card and a tripod. In the proposed framework, this scenario serves as an intermediate representation that logically links a primary item and its complements, which are difficult to capture from co-occurrence data. Large language models (LLMs) are well suited for generating such scenarios, because they can draw on extensive world knowledge to reason about real-world item use. As a first step toward a scenario-based recommendation framework, we tested the core hypothesis that LLMs can generate plausible usage scenarios using minimal input. Specifically, we prompted an LLM with only an item's category name and asked it to generate scenarios.

## 2 Related Work

Research on complementary recommendations can be categorized into four approaches: history-based [5], content-based, graph-based [4, 6, 7], and generative-model-based [4, 8, 9, 10]. History-based methods infer relationships from behavioral data such as purchase and browse logs [5]. Although they are effective in capturing real-world associations, they suffer from cold-start problems and noisy data. McAuley et al. [6] used topics from product reviews to predict relationships, and Hao et al. [7] converted textual descriptions and relationship graphs into embeddings for a graph attention network. Recent studies have explored the use of LLM, which can leverage world knowledge and commonsense reasoning to generate context-aware and interpretable recommendations without relying heavily on historical data. For instance, Huang et al. [10] proposed complementary concept generation in which a language model produces lists of related concepts for tasks, such as query suggestions. However, this was a concept-level instead of a product-level recommendation. Building on this line of work, we hypothesized that users select complements by first envisioning a usage scenario and then investigated whether LLMs can generate such scenarios as a basis for a new complementary recommendation approach.

## 3 Item Usage Scenario Generation

This section describes the design of prompts for generating realistic item-usage scenarios using an LLM. The process has three components: (1) selecting the input format, (2) crafting prompts, and (3) defining a case study to assess the scenario plausibility.

### 3.1 Input format

Generating scenarios for every product is impractical given the vast number of e-commerce items. Instead, we use product categories as scalable alternatives. Product categories are finite, standardized, and easy to manage. In this study, we used product category data from ASKUL (https://www.askul.co.jp), a major Japanese e-commerce platform. These data have a four-level hierarchical structure that provides rich item information for the LLM.

### 3.2 Prompt design

In this study, we used GPT-4o-mini to generate item usage scenarios with all the prompts and responses in Japanese. The temperature was set to 0.6 to balance creativity and coherence. The prompts include four elements that provide sufficient context and guidance.

**Instructions:** A brief explanation of the data format and task.: "*Provide information on product categories. The categories are listed from left to right in order of increasing detail. Please answer the questions accurately and specifically.*"

**Question:** A clear directive for the task: "*Please list as many specific scenarios as possible for using the specified category.*"

**Leaf Category Name:** The specific name of the target category (e.g., "*Box Tissues*").

**Full Hierarchical Path:** The complete path (e.g., "*Household Goods / Kitchenware > Tissues / Toilet Paper > Paper Towels / Daily Necessities > Tissues > Box Tissues*").

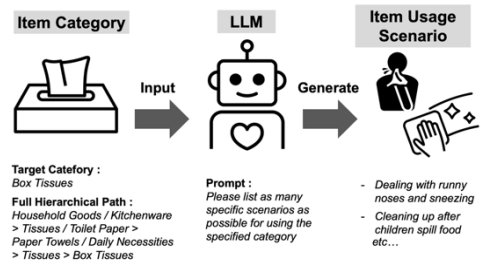

Figure 1: Overview of the item usage scenario generation process.

### 3.3 Case study for scenario evaluation

For the initial validation, we conducted a case study in the top-level category *Household Goods / Kitchenware*. Items in this category are familiar to the annotators, enabling a reliable evaluation of scenario plausibility, unlike specialized industrial products. From the 1,221 leaf categories under *Household Goods / Kitchenware*, we randomly sampled 300 categories for which the LLM generated usage scenarios. Figure 1 presents the item usage scenario generation process.

## 4 Evaluation Experiment

We conducted a human evaluation using Google Forms with 15 participants (2 faculty members and 13 undergrad / grad students) to evaluate the plausibility of the item usage scenarios. Participants were randomly divided into five groups, each assigned to 60 product categories. On average, 10 scenarios were generated per category, resulting in 2,925 scenarios across 300 categories. Each scenario was evaluated for plausibility by three participants in the same group.

The evaluation used two criteria: (1) whether the scenario was realistic and plausible and (2) whether the product category was used appropriately within the scenario. Annotators were shown two examples for "oil-based marker" to clarify the evaluation criteria: making signs or decorations for events (appropriate) and writing on a whiteboard (inappropriate), along with an explanation of why the latter is not a suitable use. The evaluation form displayed the hierarchical structure of each assigned product category and approximately ten generated usage scenarios. Participants marked the scenarios that they considered unlikely.

## 5 Result

Table 1 lists the number of LLM-generated scenarios determined as implausible by the annotators. Of the 2,925 scenarios evaluated, 2,426 (82.9%) were rated as plausible by all three annotators, whereas only 89 scenarios (3.0%) were determined as implausible by at least two annotators. These results suggest that the LLM can generally generate plausible item usage scenarios from category information alone.

| Implausible Votes | Number of Scenarios | Percentage |
|---|---|---|
| 0 | 2,426 | 82.9% |
| 1 | 410 | 14.0% |
| 2 | 82 | 2.8% |
| 3 | 7 | 0.2% |

Table 1: Evaluation results of the plausibility of item usage scenarios

Although the results support the feasibility of the proposed approach, we also analyzed the variability in annotators' judgments to assess the subjectivity of scenario plausibility. Differences among individuals were substantial; on average, annotators marked 6.78% of the scenarios as implausible, with a high standard deviation of 4.61 and, ranging from a minimum of 2.38% to a maximum of 21.5%. This variance indicates that the perceptions of plausibility are highly subjective. This finding reinforces our initial claim that complementary relationships are inherently ambiguous, and highlights a key challenge for future research: developing recommender systems and evaluation frameworks that account for human subjectivity.

## 6 Discussion

### 6.1 Comparing of plausibility and implausibility judgments

The high plausibility rate of 82.9% likely reflects the ability of the model to generate scenarios that mirror real consumer behavior. For instance, scenarios for the lap blanket category included direct uses such as "counteracting office air conditioning" as well as purchase motivations such as "giving as a gift." Such realistic situations are likely to have contributed to the high scores. Conversely, 89 scenarios determined implausible by two or more evaluators were manually reviewed and classified into five main error categories:

**Inappropriate use (35 cases):** The most frequent error type, accounting for 39.3% of the implausible scenarios. This category applies when the proposed use contradicts the intended function, properties, or common-sense usage of the item. For instance, kitchen duckboards and mats for "draining food ingredients in commercial kitchens are both unsanitary and unintended.

**Not an item usage scenario (21 cases):** The second most common error type, covering texts that describe actions related to an item, such as selection, purchase, or maintenance, instead of its actual use. Examples include "selecting products specially made for babies with allergies or sensitive skin" for a baby soap category, and "scheduling regular veterinary visits for a guinea pig's health" for a pet food category. Although relevant, they do not depict the item's direct functional use. In this study, usage scenario refers to a situation in which an item is directly and functionally involved in an action or task.

**Suboptimal item for the scenario (15 cases):** This error type involves a poor match between the item function and the proposed use. For instance, using a plastic shopping bag "for temporary document storage at school or work" misaligns the features of the item with the intended application.

**Use is indirect, auxiliary, or excessively abstract (10 cases):** This category included scenarios where the item was technically used but not central to the task. For instance, a pot lid described as "for efficient use when cooking large quantities at once" plays only a secondary role in the main task.

**Unrealistic scenario (8 cases):** A small number of scenarios involved highly improbable situations. For instance, "hosting a self-hair-removal event at home" using hair removal products is rarely practiced in the real world.

### 6.2 Detailed analysis of implausible scenarios

In this section, we analyze in detail the scenarios determined as implausible by two or more annotators. Of the 300 categories, 89 errors occurred in 52. The distribution was long-tailed; in the most affected categories (45 of 52), only one or two implausible scenarios appeared, indicating that the output of the LLM was generally reliable. However, a few categories accounted for a disproportionate share of the errors. The *kitchen paper holder* category contained seven implausible scenarios. A manual review of these cases revealed a specific failure pattern. In some instances, the name of a leaf category (e.g., *kitchen paper holder*) contained the name of a more commonly used item (*kitchen paper*) that also appeared in the upper-level category. This overlap could cause the LLM to misinterpret the prompt and instead generate scenarios intended for a more common item. In this case, numerous scenarios described the use of kitchen paper (e.g., wiping up spills), even though the prompt clearly referred to the holder.

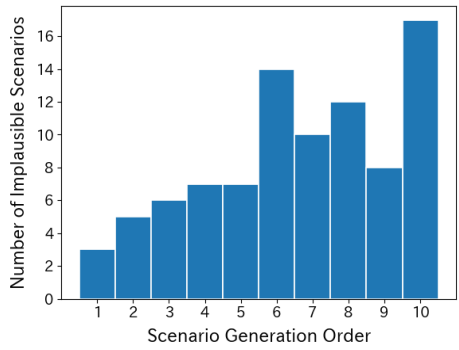

Figure 2: Number of implausible scenarios by generation order.

Next, to explore strategies for reducing these errors, we examined when implausible scenarios tended to occur in the generation sequence. This analysis was motivated by prior findings that LLMs often exhibit positional bias, producing higher-quality outputs earlier in a sequence. As shown in Figure 1, our results support this pattern: implausible scenarios appeared significantly more often in later positions.

This finding has a clear practical value. Limiting the selection to the first five scenarios generated per category reduced the number of implausible cases from 89 to 28, thereby eliminating the most problematic cases. This filtering also reduced the number of categories containing implausible scenarios from 52 to 19. These results suggest that using only the top-ranked scenarios is a simple yet effective way to improve the quality and reliability of the generated data.

## 7 Conclusion

In this study, we investigated the plausibility of LLM-generated item usage scenarios for complementary recommendation. Manual annotation revealed that approximately 85% of scenarios were plausible, with implausible cases often stemming from item misinterpretation or contextual errors. We also found that earlier-generated scenarios were more reliable, suggesting a simple filtering method. Unlike conventional methods based on historical co-occurrence, our scenario-based approach captures the underlying context of item use. This offers a path toward more explainable, context-aware recommendations better equipped to handle the cold-start problem.

Future work will focus on automatically extracting complementary item pairs from scenarios and integrating them into practical systems. We will explore strategies such as enriching candidate generation for existing models and using scenarios as explainable justifications to build user trust. We also plan to expand our evaluation to a broader range of categories, including specialized domains.

## Acknowledgement

This work was supported by JSPS KAKENHI, Grant Numbers JP23K21724, JP23K24953, JP24K21410.